\documentclass[%
 aip,
rsi,%
 amsmath,amssymb,
 reprint,%
 floatfix,
]{revtex4-1}

\usepackage{color}
\usepackage{graphicx}% Include figure files
\usepackage{dcolumn}% Align table columns on decimal point
\usepackage{bm}% bold math
\usepackage{multirow}
  \usepackage{algorithm} %sagar
  \usepackage{algpseudocode}  %sagar
\newcommand{\metaz}{\overline{\mathbf z}} 

\newcommand{\rxn}[2]{{\bf #1} $\rightarrow$ {\bf #2}}
\begin{document}

\preprint{AIP/123-QED}

%\title{{\it Ab Initio} MD Simulations with Hybrid Functionals: Implementation and Application}
\title{Mean Force Based Temperature Accelerated Sliced Sampling: Efficient
Reconstruction of High Dimensional Free Energy Landscapes
%{\it Ab Initio} Molecular Dynamics Simulations of Chemical Reactions with Hybrid Functionals and Plane Wave Basis Set
}
%\title[Sample title]{Sample Title New:\\with Forced Linebreak\footnote{Error!}}% Force line breaks with \\
%\thanks{Footnote to title of article.}

\author{Asit Pal}
\affiliation{ 
Department of Chemistry, Indian Institute of Technology Kanpur, Kanpur - 208016, India}
%\\This line break forced with \textbackslash\textbackslash

\author{Subhendu Pal}
\affiliation{ 
Department of Chemistry, Indian Institute of Technology Kanpur, Kanpur - 208016, India}%\\This line break forced with \textbackslash\textbackslash

\author{Shivani Verma}
\affiliation{ 
Department of Chemistry, Indian Institute of Technology Kanpur, Kanpur - 208016, India%\\This line break forced with \textbackslash\textbackslash
}%

\author{Motoyuki Shiga}
\email{shiga.motoyuki@jaea.go.jp}
\affiliation{ 
Center for Computational Science and E-Systems,Japan Atomic Energy Agency, 148-4-4 Wakashiba, Kashiwa, Chiba, 277-0871, Japan
}%

\author{Nisanth N. Nair}
\email{nnair@iitk.ac.in}
\affiliation{ 
Department of Chemistry, Indian Institute of Technology Kanpur, Kanpur - 208016, India%\\This line break forced with \textbackslash\textbackslash
}%

 %\homepage{http://www.Second.institution.edu/~Charlie.Author.}
%\affiliation{%
%Second institution and/or address%\\This line break forced% with \\
%}%

\date{\today}% It is always \today, today,

\begin{abstract}
{Temperature Accelerated Sliced Sampling (TASS) is an efficient method to compute high dimensional free energy landscapes. 
The original TASS method employs the Weighted Histogram Analysis Method (WHAM) which is an iterative post-processing to reweight and stitch high dimensional probability distributions in sliced windows that are obtained in the presence of restraining biases.
The WHAM necessitates that TASS windows lie close to each other for proper overlap of distributions and span the collective variable space of interest. 
On the other hand, increase in number of TASS windows implies more number of simulations, and thus it affects the efficiency of the method.
To overcome this problem, we propose herein a new mean-force (MF) based reweighting scheme called TASS-MF, which enables accurate computation with a fewer number of windows devoid of the WHAM post-processing.
Application of the technique is demonstrated for alanine di- and tripeptides {\it{in vacuo}} to compute their two- and four-dimensional free energy landscapes, the latter of which is formidable in conventional umbrella sampling and metadynamics. 
The landscapes are computed within a kcal~mol$^{-1}$ accuracy, ensuring a safe usage for broad applications in computational chemistry.}
%Computational overhead can be substantially decreased by neglecting some of the 
% orbital products based on the extent of their contribution. % to exact exchange energy.
%
%For this, 

%explicit water model.
%employing Well-Sliced Metadynamics approach.
%
%
%Valid PACS numbers may be entered using the \verb+\pacs{#1}+ command.
\end{abstract}

\maketitle

\section{\label{sec:intro}Introduction}
%
% time scale bottleneck of normal MD
Free energy barriers, and free energy difference between  
reactants  and products,  are the two thermodynamic 
quantities of interest in predicting the 
spontaneity and kinetics of the chemical
reactions and other physio-chemical transformations.
In this respect, computing free energy landscape of such processes as a function of few collective variables (CVs) is a commonly employed strategy.\cite{Peters:Book,Tuckerman:Book,Vanden:2009:jcc,vanGunsteren:JCC:Rev:2010,Ciccotti:12,Pratyush:2016,shalini:review:2019,vashisth:review:2019}
Molecular Dynamics (MD) combined with enhanced sampling techniques are widely used for this purpose.
Enhanced sampling methods are essential to accelerate the 
transitions from one free energy basin to another. 
%
%computational community to explore the free energy landscapes of several processes in order to study the feasibility of these processes. Now for a canonical ensemble, the probability of observing a particular configuration R having the potential energy $U$ is proportional to $\exp{[-\beta U}$(R)] where {$\beta = (k_{B}T)^{-1}$}, $k_{B}$ is the Boltzmann constant and $T$ is the temperature. So MD can be proven to be inefficient for the processes having several local minima separated by a high free energy barrier due to the fact that time scales accessible by integrating equations of motion of atoms is much less than the time scale of the process of our interest. These are called rare events.\\ 
%
Accelerating CVs by bias potentials\cite{us:orig,Darve:2001,Laio:PNAS:02,Naidoo:2009} and high-temperature\cite{Rosso:JCP:2002,Tuckerman:Book,TAMD:1,Tuckerman:2008} are some of strategies proposed for this purpose.
%that are certain functions of atomic coordinates.\cite{TODO}
%
%Correct choice of CVs is critical in obtaining correct free energetics
%for such methods.
%
%Other than the requirement that these coordinates should characterize the reactant, product and the transition states,
%they should also encompass transverse coordinates that are essential for a thorough conformational sampling.
%
Alternative approaches include accelerating the system dynamics by flattening the underlying potential energy landscape by
bias potentials, replica--exchange based global-tempering and other generalized ensemble methods.\cite{Tuckerman:Book,Voter:1997,Roman:2006,Berne:2005,ITS,Grubmuller:PRE:95,McCammon:2004,Okamoto:2001,Okumura:2012,Morishita:2012,Kasahara:2019} %TODO-shivani: cite Okamoto, Okumura, Kasahara, 

One of the simplest approaches that employ biased sampling of CVs is Umbrella Sampling (US).\cite{us:orig,Kastner:11}
Here a bias potential,
\begin{equation}
 W_h^{\rm b}(s) = \frac{1}{2} k \left [ s(\mathbf R) - \xi_h \right ]^2 , \enspace h=1,\cdots,M \enspace , 
 \label{us:bias}
 \end{equation}
is applied to restrain the CV $s(\mathbf R)$ at $M$ discrete values $\{ \xi_h \}$.
In the above, $\mathbf R$ is the set of atomic coordinates, and $k$ is the coupling constant.
Biased canonical probability distribution of $s$ obtained for each umbrella window 
is given as,
\begin{eqnarray} 
\tilde P_h(s^\prime) &=& \frac{1}{\tilde Z} \int d \mathbf R \,  e^{ - \beta \left [ U(\mathbf R) + W_h^{\rm b}(s) \right ]} \, \delta \left ( s(\mathbf R) - s^\prime \right ) \nonumber \\[1ex]
&\equiv& \left < \delta \left ( s(\mathbf R) - s^\prime \right ) \right >_{ W_h^{\rm b}} 
\end{eqnarray}
where 
\[ \tilde Z = \int d \mathbf R \,  e^{ - \beta \left [ U(\mathbf R) + W_h^{\rm b}(s) \right ] } \enspace ,  \]
$U$ is the potential energy,  $\beta = \left ( k_{\rm B} T \right )^{-1}$, 
$k_{\rm B}$ is the Boltzmann constant, and $T$ is the temperature.
In the above $\left <\cdots \right >_{W_h^{\rm b}}$ specifies the 
ensemble average in the presence of the bias $W_h^{\rm b}$.
The Weighted Histogram Analysis Method (WHAM)\cite{wham:1,wham:2} is then employed to combine the biased probability distributions and reweight them to obtain final (unbiased) distribution $P(s)$.
In WHAM, this is done by computing 
\begin{equation}
\label{wham}
P(s)=\frac{\sum_{h=1}^{M}n_{h}{\tilde{P}_{h}(s)}}{\sum_{h=1}^{M}n_{h} g_h \exp{[-\tilde{\beta}W_{h}^{b}(s)]}}    
\end{equation}
where, 
\begin{equation}
g_h^{-1} = \int{ds \exp{[-\tilde{\beta}W_{h}^{\rm b}(s)]}} \, {P}(s)     \nonumber
\end{equation}
is unknown. 
An iterative procedure is employed, where the iteration begins with $g_h=1$, and improved at every step based on the
$P(s)$ computed in the previous step.
For the correct convergence of $P(s)$, a proper overlap of $\tilde P_h(s)$ with its neighboring distributions  is essential.
It is to be noted that the width of the distribution $\tilde P_h(s)$ depends on the value of $k$; Higher the value of $k$, more accurate will be the free energy estimate.
On the other hand, increasing the value of $k$ will make the distribution narrower, and thereby the extent of overlap between the neighboring distributions will become smaller.
As a result, more umbrella windows have to be used with high values of $k$
for obtaining accurate results.
Further, when dealing with high dimensional CV-space, 
long MD simulations are required to achieve sufficient overlap of distributions.
%
%{\tt \color{red} <<<TILL HERE>>>}
Due to this limitation, US is often used to explore only a one-dimensional or a part of a two-dimensional CV-space.

%Here the computational cost increases exponentially with the number of CVs which makes this technique limited to only few CVs. \\

% Importance of accelerated sampling techniques.
%In order to sample such rare events properly within feasible simulation time, we have to either  modify the potential energy, $U$(R) to a lower value or we have to modify the Boltzmann factor $\beta$. 
%

%An improvement over US method was recently proposed.
Several techniques were proposed to improve the limitations of US.\cite{Karplus:US:64,Kroon:JCP:1992,Mezei:JCP:87,Berneche:JCP:2013,Shalini:2017,UI,Kastner:11} 
Temperature Accelerated Sliced Sampling (TASS) method~\cite{Shalini:2017,shalini:review:2019}
extends the US method to high 
dimensions by combining it with 
temperature accelerated molecular dynamics (TAMD)/driven adiabatic 
free energy dynamics (d-AFED)~\cite{AFED2,TAMD:1,Tuckerman:2008} 
and metadynamics~\cite{Laio:PNAS:02} %TODO
methods.
TASS uses the extended Lagrangian,
\begin{eqnarray}
\label{lag:tass}
    \mathcal L_h({\bf R,\dot{R},z,\dot{z})} &=&  \mathcal L^{0}({\bf R,\dot{R}})+ \nonumber \\
   & & \sum_{\alpha = 1}^{n}\left [\frac{1}{2}{\mu_{\alpha}}{{\bf \dot{z}}_{\alpha}}^2+\frac{\kappa_{\alpha}}{2}(s_{\alpha}({\bf R})-z_{\alpha})^2\right] -  \nonumber \\
   & & W_{h}^{\rm b}(z_1)- V^{\rm b}(\metaz,t),  \nonumber \\
   & & h=1,\cdots,M,
   %+{\text{bath}}({\bf P};T)+{\text{bath}}({\bf p};\Tilde{T})
\end{eqnarray}
where $\mathcal L^0$ is the original Lagrangian of the system,
$\mu_\alpha$ is the mass of the auxiliary degree of freedom $z_\alpha$, 
and $\kappa_\alpha$ is the restraining force of the spring that couples
the $z_\alpha$ and the corresponding CV $s_\alpha$.
 Two kinds of bias potentials $W_h^{\rm b}(z_1)$ and
$V^{\rm b}(\metaz,t)$, are added on the auxiliary degrees of freedom. 
The bias $W_h^{\rm b}(z_1)$ is the umbrella bias potential given by
Eqn.~\ref{us:bias}, acting on the auxiliary variable $z_1$.
A metadynamics bias,\cite{Laio:PNAS:02,Iannuzzi:03} $V^{\rm b}(\metaz,t)$ 
is applied along a small set of auxiliary variables $\metaz \equiv (z_2,\cdots,z_m)$, 
and $m\leq n$.
The dimension of the  auxiliary vector space ${\metaz}$ is less than or equal to that of ${\bf z}$.
It is preferred to choose
\begin{equation}
\label{vbias:mtd}
    V^{\rm b}(\metaz,t) = \sum_{\tau<t} w_\tau\,\exp{\left[-\frac{{\left\{\metaz -  \metaz_\tau \right\}}^2}{2(\delta z)^2}\right]}
\end{equation}
with
\begin{equation}
 w_\tau = \omega_0\exp\left[{-\frac{V^{b}( \metaz_\tau , \tau)}{k_B\Delta T}}\right],  
 \end{equation}
and $\metaz_\tau \equiv \metaz(\tau)$ 
as in well-tempered metadynamics (WT-MTD).\cite{mtd:well:08,Voth:14}
In the above, $w_\tau$ is the height of the Gaussian deposited at time $\tau$, $\delta z$ is the width of the Gaussian, and $\Delta T$ is a parameter.
In TASS, the auxiliary variables are thermostatted to $\tilde T$~K,  while 
the physical system is thermostatted to $T$~K, with $\tilde T \gg T$.
Here we define, 
$\tilde \beta = \left ( k_{\rm B} \tilde T \right )^{-1}$ and
$\beta = \left ( k_{\rm B} T \right )^{-1}$.
%
%The use of metadynamics bias is optional. 
%
The probability distribution
for each TASS window $h$ is first calculated as
\begin{equation}
\label{e:mtd:reweight}
\tilde P_h(\mathbf z^\prime) = \frac{ \int dt \, A_h(t) \, \prod_\alpha^n \, 
\delta \left (z_\alpha(t) - z^\prime_\alpha \right ) }
{\int dt \, A_h(t) }
\end{equation}
where 
\begin{equation}
\label{a_tau}
A_h(t) = \exp \left [ 
         \tilde \beta 
    \left \{  
       V_h^{\rm b}\left (\metaz_t, t \right ) 
        - c_h(t) 
   \right  \} 
    \right ]
\end{equation}
and
\begin{equation}
c(t) = \frac{1}{\tilde \beta} \ln 
\left [ \frac{\int d \metaz \, \exp \left [ \tilde \beta \gamma V^{\rm b}(\metaz,t) \right ]
}
{\int d \metaz \, 
\exp \left [ \tilde \beta (\gamma - 1) V^{\rm b}(\metaz,t) \right ]
}
\right ] \enspace . 
\end{equation}
As next, high dimensional probability distribution $\tilde P(\mathbf z)$
is  obtained by Eqn.~\ref{wham}.
Free energy surface at temperature $T$ is obtained as
\begin{eqnarray}
\label{fes:tass}
F(\mathbf z) = - \frac{1}{\tilde \beta} \ln \tilde P(\mathbf z) 
\end{eqnarray}
as in TAMD/d-AFED.
The main advantage of TASS is that a large number of collective variables can be
used by virtue of the temperature acceleration of the auxiliary space.
Umbrella bias provides a way to achieve a directed or controlled sampling
when used along an appropriate CV.
The method also permits to use different transverse coordinates for different
umbrella windows depending on the requirement.
This is possible as free energy slices are independently computed in the corresponding
windows.
%
%If $n$ CVs are extended to $n^\prime$ CVs, in a particular umbrella window, 
%then $n$ dimensional distribution is obtained by appropriate projection
%of the probability density $\tilde P_h^\prime$ in the high-dimensional space:
%\begin{eqnarray}
%\tilde P_h(\mathbf z) &=& \int d z_{n+1} \cdots d z_{n^\prime}   \nonumber \\
%& & \, \tilde P_h^\prime \left ( z_1,\cdots,z_n, z_{n+1},\cdots,z_{n^\prime} \right )
%\enspace .
%\nonumber
%\end{eqnarray}
%
%We emphasize at this point that when $W_h^{\rm b}$ and $V^{\rm b}$ are zero,
%TASS is identical to the TAMD/d-AFED method.
%
%The crucial component of the TASS method is thus the differential bias acting on the CVs.

However, reconstruction of free energy surfaces by WHAM encounters problems
when dealing with large number of CVs for the reasons discussed earlier.
In this paper, we propose a way to reconstruct high dimensional 
free energy surfaces in TASS simulations using a mean force based method, and 
dispense the computationally
inefficient WHAM approach.
Most importantly, the proposed method does not involve an iterative scheme, unlike WHAM.  
The new method is quite general, and is applicable not only to TASS, but also to US and WT-MTD simulations.
%
%It is also preferable to use the optimal number of umbrella windows, depending
%on the nature of the free energy surface.
%%Advantage of TASS
%%Issues with TASS.

%
%TODO-MS: Do we have to add any references here? Mean-force papers from MET?
%
%We extend the concept in the reconstruction of high-dimensional free energy
%surface using the extended Lagrangian formalism and with different biases acting along
%different CVs, and high temperature felt on the extended variables.

%
%
%The Temperature Accelerated Molecular Dynamics (TAMD)/driven Adiabatic Free Energy Dynamics (d-AFED)\cite{Tuckerman:Book,TAMD:1,Tuckerman:2008}, and
%Temperature Accelerated Sliced Sampling (TASS) are few other methods
%which take the advantage
%of auxiliary variables coupled to the CVs.
%

\section{Theory}

% Our Present Work --------------
%Due to the slow and costly WHAM, the reweighing of TASS becomes computationally expensive. In %our present work we propose a new reweighing scheme for TASS.
%
%Here we rearrange the free energy in such a way that it gets separated into two parts. One part corresponds to the free energy component only along Umbrella CV and the other part corresponds to the free energy along other orthogonal CVs for a fixed value of Umbrella CV. Then we calculate the free energy along Umbrella CV by integrating the mean force acting on this CV. 

Mean force based computation of free energies is core to several techniques.\cite{ciccotti:89,ciccotti:98,McCammon:JCP:2008,Darve:2001,Tuckerman:2012,Morishita:JCTC:2017}
Mean force based reconstruction of free energy surfaces in the frame work of  
extended Lagrangian has been reported earlier for adaptive bias sampling (ABF) \cite{Darve:Book,Darve:2001,darve:jcp:2008,Gabriel_2007,Chipot:2015,Henin:JPCB:2017}, logarithmic mean-force dynamics\cite{Morishita:2012} %TODO-shivani (Tetsuya Morishita's paper here)
TAMD/d-AFED, UFED and metadynamics.\cite{Tuckerman:2012,Tuckerman:JCTC_2014}
Such techniques use the gradient of the underlying free energy to compute free energy surfaces using thermodynamic integration as
\begin{eqnarray}
F(z) = \int^{z} d z^\prime \left < \frac{dF}
{d z^\prime} \right >  \enspace .
\end{eqnarray}
Such integration become tricky with increasing dimensions.
Using a basis set expansion\cite{Tuckerman:2012} and artificial neural network representation\cite{Tuckerman:PRL:2017}
of free energy will be more efficient approaches to overcome such issues to some extent.
Here we present an alternative method that suits the TASS approach, and is efficient for dealing with high dimensions.
We will utilize the advantages of employing mean-forces  to combine independently sampled probability distributions (or free energy slices) by the TASS Lagrangian (Eqn.~\ref{lag:tass}) to reconstruct high-dimensional free energy surfaces.
The challenge here is to reconstruct a high-dimensional free energy landscape where the
CVs are differently biased. 

First, we write the derivative of
projected free energy, $F_1(z_1)$, along the auxiliary variable $z_1$ in which the umbrella bias is active.
\begin{equation}
F_1(z_1) = - \frac{1}{\tilde \beta} \ln P_1(z_1)
\end{equation}
where,
\begin{equation}
    P_1(z_1)=\int dz_2 \cdots dz_n \, e^{- \tilde  \beta F(z_1,\cdots,z_n)}
\end{equation}
%along which the US bias is applied, at $z_1=\xi_h$, 
%as,
%where $\frac{\partial F}{\partial s_1}$ is the partial derivative of free energy with respect to $s_1$ and can be written as \cite{darve:jcp,Darve:Book,darve:jcp:2008}
Since US bias is applied along $z_1$, the force at $z_1=\xi_h$ is given by
\begin{equation}
\left ( \frac{d F_1}{d z_1} \right)_{\xi_h} = 
-{\left\langle  
k_{h} \left [ z_1 - \xi_h \right ]
\right\rangle}_{\xi_h}, \enspace h=1,\cdots,M.
\end{equation}
%%where $f$ is the instantaneous force on the variable $s_1$. So, derivative of free energy is the %negative of mean force acting on $s_1$ under the condition $s_1=s_1^\prime$. For our case, umbrella %potential, $W_h^b$, which is expressed through Eq.(1), is applied along $s_1$. So, the mean force on %$s_1$ can be written as follows,
%The right hand side of the above expression is the
%mean force computed for $z_1=\xi_h$, i.e. the
%equilibrium position of the umbrella bias, and 
%the index $h=1,\cdots,M$ runs over all the umbrella
%windows.
%
Note that, other approach could be to compute the mean-force on real CVs (i.e., $\mathbf s$, and not the auxiliary coordinates, $\mathbf z$), as in Ref.~\cite{Tuckerman:2012}.
Since a subset of the auxiliary coordinates, $\metaz$,
is biased by time dependent metadynamics potential in the TASS formalism,
we can relate,
\begin{widetext}
\begin{eqnarray}
{\left\langle  
k_{h} \left [ z_1 - \xi_h \right ]
\right\rangle}_{\xi_h}
=  
\frac{
\int{d \mathbf z \; 
\left [k \left (z_1-\xi_h^h \right ) \right ]\;
\exp \left [  
\tilde{\beta}\{V^{\rm b}(\metaz,t)-c (t) \}
\right ]
}
}{
\int{ d\mathbf z \; \exp \left [  
\tilde{\beta}\{V^{\rm b}(\metaz,t)-c(t) \}
\right ] }
}
%
%\int{dz_2 \cdots dz_n\; \left [k \left (z_1-\xi_1^h %\right ) \right ]\; 
%\exp{ \left [\tilde{\beta}{V_h^b(s_2,t)-c_h(t) } 
%\right ]
%\} 
%}{\int{ds_2 ...ds_n\; \exp{[\tilde{\beta}\{V_h^b(s_2,t)-c_h(t)\}]} }} 
\end{eqnarray}
\end{widetext}
which in turn can be computed by the time average,
\begin{eqnarray}
\frac{ \int dt \, A(t) \, 
\left [k \left (z_1-\xi_h^h \right ) \right ] }
{\int dt \, A(t) }
\end{eqnarray}
and the time dependent reweighting factor $A(t)$ is given by Eqn.~\ref{a_tau}.
Employing these equations, we can compute projected free energy $F_1$
along $z_1$ from a TASS simulation as follows:
\begin{eqnarray}
\label{fes1d}
F_1(z_1) &=& \int^{z_1} 
dz_1^\prime    
 \, \left ( \frac{d F_1 }
{d z_1^\prime} \right ) \nonumber \\
&\approx & - \sum_{h=1}^{M^\prime-1} \Delta \xi_h \, \, w_h 
\left [ g_h  + g_{h+1} \right ] \, \,
\end{eqnarray}
where
\begin{eqnarray}
g_h=\langle k_h \left [ z_1 - \xi_h \right ]  \rangle_{\xi_h} \enspace .
\end{eqnarray}
Here, $M^\prime$ is the grid index corresponding to the CV value $z_1$.
This integral can be computed in a straight forward manner 
using the trapezoidal rule, with $\Delta \xi_h = \xi_{h+1} - \xi_{h}$, and the integration weights $w_h= \frac{1}{2}$ for all the
values of $h$.
We stress that the above equation reconstructs only the projection
of the free energy surface $F(\mathbf z)$ along the US coordinate $z_1$, i.e. $F_1(z_1)$.
Now, the multidimensional free energy surface can related to $F_1(z_1)$ as follows:
\begin{eqnarray} 
F(z_1,\cdots,z_n) &=& 
- \frac{1}{\tilde  \beta} \ln \, \left [
e^{-\tilde  \beta F(z_1,\cdots,z_n)} \right ]  
\nonumber \\
&=& - \frac{1}{\tilde  \beta} \ln \left [ \frac{P_1(z_1)}{P_1(z_1)}
e^{-\tilde  \beta F(z_1,\cdots,z_n)} \right ]  
\nonumber \\
&=& - \frac{1}{\tilde  \beta} \ln P_1(z_1) - 
\frac{1}{\tilde  \beta} \ln \frac{e^{- \tilde  \beta F(z_1,\cdots,z_n)} }{P_1(z_1)} 
\nonumber \\
&=& F_1(z_1) - \frac{1}{\tilde  \beta}
\ln \frac{e^{- \tilde  \beta F(z_1,\cdots,z_n)}}{
\int dz_2 \cdots dz_n \, e^{- \tilde  \beta F(z_1,\cdots,z_n)}} \nonumber \\
&=& F_1(z_1) - \frac{1}{\tilde \beta} \ln \tilde P_{z_1}
(z_2,\cdots,z_n) \nonumber \\
&=& F_1(z_1) + \Delta F_{z_1}(z_2,\cdots,z_n) 
\label{fe:correction}
\end{eqnarray}
where,
\begin{equation}
\label{fes:corr2}
   \Delta F_{z_1}(z_2,\cdots,z_n) =
   -\frac{1}{\tilde{\beta}} \ln \tilde P_{z_1}(z_2,\cdots,z_n )
\end{equation}
is the $n-1$ dimensional slice of the free energy surface at $z_1$, and $\tilde P_{z_1}$ is the corresponding slice of the probability
distribution obtained at temperature $\tilde T$~K.
We can use the relationship in Eqn.~\ref{fe:correction} to reconstruct the high dimensional free energy landscape, $F(\mathbf z) \equiv F(z_1,z_2,\cdots,z_n)$.
%
%The distribution $\tilde P_{z_1}(z_2,\cdots,z_n)$ needs to be co binning $\mathbf z$ along TASS trajectory at umbrella window $h$ and $\xi_h \equiv z_1$.
%
Since in TASS, time dependent bias  
acts along a subset of the orthogonal CV space $z_2,\cdots,z_n$, the reweighting approach as discussed in Eqn.~\ref{e:mtd:reweight}
has to be used for obtaining unbiased distribution $\tilde P_{z_1}(z_2,\cdots,z_n)$ by binning.
%   \left[\frac{\exp{\{-\tilde{\beta}F(s)\}}}{\int ds^\prime \exp{\{-\tilde{\beta}F(s)\}}} \right]}\\[3mm]
%                      % & = F(S_1,S_2) + \frac{1}{\beta}\ln{\left\{\int dS_2\exp{\{-\beta F(S_1,S_2)\}}\right\}}\\[3mm]
%                     & = F(s) - F(s_1)
%\end{align}\\
Interpolation schemes can be used to obtain free energy 
values between the grid points  $\xi_1,\cdots,\xi_M$ along $z_1$.

The procedure explained here does not require WHAM
for stitching the free energy slices together to obtain
 high dimensional free energy surfaces.
This approach which uses the mean-force will be denoted as TASS-MF and the WHAM based approach to reconstruct the TASS free energy surfaces will be called TASS-WHAM, hereafter.

In practice, we perform the following steps 
in reconstructing the high-dimensional free energy surface:
\begin{enumerate}
    \item Perform TASS simulations for the windows $h=1,\cdots,M$ using the 
    Lagrangian given in Eqn.~\ref{lag:tass}.
    \item Compute $\left \{ V_h^{\rm b} \right \}$ and $\left \{ c_h(t) \right \}$, for $h=1,\cdots,M$.
    \item From the trajectory of the auxiliary variables, $z_1,\cdots,z_M$, compute the one dimensional free energy profile 
    $F_1(z_1)$ using Eqn.~\ref{fes1d}.
    \item Compute the high dimensional free energy slice for 
    $z_1 \in \left ( \xi_1,\cdots,\xi_M \right ) $
    using Eqn.\ref{fes:corr2}.
    \item Reconstruct the high dimensional free energy surface employing Eqn.~\ref{fe:correction}.
\end{enumerate}

\section{Result and Discussion}
\subsection{Alanine Dipeptide in Vacuum}

\begin{figure}[htbp]
\centering
    \includegraphics[width=0.50\textwidth]{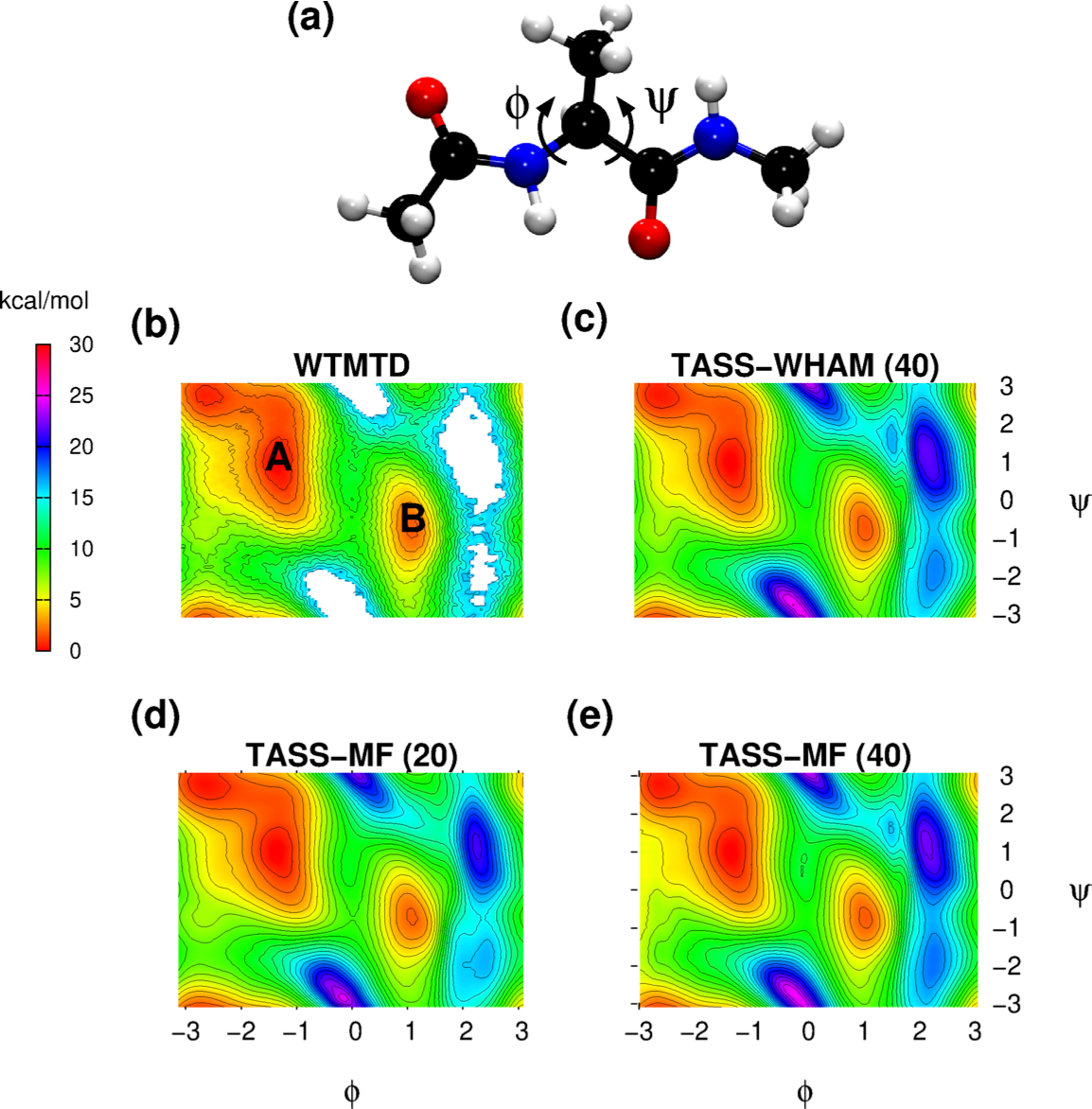}
    \caption{%{\tt \color{red} FIGURE 1 of paper}
    (a) Ball and stick representation of alanine
dipeptide. $\phi$ and $\psi$ are the Ramachandran angles, specifically the torsional angle between atoms C-N-C$_{\alpha}$-C and N-C$_{\alpha}$-C-N, respectively, as shown in the figure. Color code: H (white), C (black), O(red), and N (blue). The free energy surface reconstructed in the $(\phi,\psi)$-space computed from (b) WTMTD (100~ns), and (c) TASS-WHAM with 40 equidistant windows
are used as the reference data. The same computed using TASS-MF with 20 and 40 equidistant
windows are shown in the subfigures (d), and (e). The contours are drawn for every 1 kcal mol$^{-1}$}
    \label{fig:aladi_fes}
\end{figure}

To demonstrate the accuracy of this method, free energy surface of alanine dipeptide {\em in vacuo} as a function of Ramachandran angles ($\phi,\psi$) was computed using various methods; see Fig.~\ref{fig:aladi_fes}.
%
%Here the free energy for conformational changes of alanine dipeptide is explored as a function of two backbone angles $(\phi,\psi)$ [figure 1(a)]. 
%
The molecule was modelled by the ff14SB \cite{ff14SB} AMBER force field and MD simulations were performed by using AMBER 18\cite{amber18} patched with the PLUMED interface. \cite{plumed} 
{The same set of calculations was repeated using the PIMD code where the TASS method has been implemented.\cite{Shiga:2016,Shiga:PIMD2020} } %TODO-shivani 
The time step used for integrating the equations of motion is $1$~fs.
%Each each umbrella window,  10~ns long. 
%
%
We performed TASS simulations by applying umbrella bias along $\phi$ and WTMTD bias along $\psi$.
Three sets of TASS simulations were executed by varying number of  umbrella windows:
(a) 40; (b) 30; (c) 20.
The umbrella restraints were centered equal-distant in the domain $\phi \in ( -\pi , \pi ]$ and $k=239$~kcal~mol$^{-1}$~rad$^{-2}$ was taken. 
For the Gaussian bias (Eqn.~\ref{vbias:mtd}), we chose $w_0=0.57$~kcal~mol$^{-1}$,  
$\delta z=0.05$~rad, $\Delta T=2700$~K, and the bias was updated every 500~fs. 
We took $\kappa=1258$~kcal~mol$^{-1}$~rad$^{-2}$ and $\mu=50$~amu~{\AA}$^2$~rad$^{-2}$. 
Calculations were done at canonical ensemble with $T=300$~K, and an 
%over-damped
Langevin thermostat
having a frictional coefficient of 0.1 fs$^{-1}$ was employed to control the
temperature of the physical system.
We used $\tilde T= 1000$~K  for the extended system, and the temperature
of the extended variables was maintained using a separate 
%over-damped
massive Langevin thermostat 
with a frictional coefficient of 0.1 fs$^{-1}$.
%
%The initial structure of each window was obtained from the equilibrated structure of
%the adjacent window.
%
%{\tt \color{red} Equilibration run for each window was carried out for about 5~ns.}
%
For each window, we performed {20}~ns long production run.
%
%\begin{figure}[htbp]
%\centering
%    \includegraphics[width=0.50\textwidth]{triala_figure.pdf}
%    \caption{ \label{fes:triala} 
%    (a) Ball and stick representation of alanine
%tripeptide. 
%
%The Ramachandran angles which are used as the CVs are shown. 
%
%The free energy surfaces along 
%(a) $(\phi_1,\psi_1)$, 
%5(b) $(\phi_2,\psi_2)$,
%(c) $(\phi_1,\phi_2)$,
%computed using TAMD simulations are shown. 
%
%Free energies are in kcal mol$^{-1}$ and 
%contours are drawn for every 1 kcal mol$^{-1}$.
%
%For reference, the free energy surface computed
%as a function of $(\phi_1,\phi_2)$ using WHAM reweighting 
%is also shown (d).
%}
%    \label{fig:conf}
%\end{figure}
%
As reference free energy data, we performed a conventional WTMTD simulation for 100~ns, where the coordinates $\phi$ and $\psi$ were biased using a two-dimensional Gaussian bias-potential (as in Eqn.~\ref{vbias:mtd}).
We used $\Delta T=1500$~K, otherwise all the metadynamics parameters were the same as that in the TASS simulations.

In Figure~\ref{fig:aladi_fes}(b-e), we have presented the free energy surfaces using WTMTD, TASS with WHAM based reweighting with equidistant 40 umbrella windows (TASS-WHAM (40)), and the new method presented here, i.e., using the mean-force (TASS-MF).
TASS-MF reweighting was carried out with 40, 30 and 20 equidistant windows, and they are indicated as TASS-MF (40), TASS-MF (30), and TASS-MF (20), respectively.
Clearly, the positions of the minima and saddles are exactly identical in all the landscapes;
See Fig~\ref{fig:aladi_fes}(b)-(e).
%{\color{red}SI~Table~XX}. %TODO. Add a supporting information with tables of the location of minima computed from BMTASS and WTMTD.
%
A quantitative comparison of the free energy barriers
on the $F(\phi,\psi)$ surface was carried out for WTMTD, TASS-WHAM, TASS-MF simulations; 
see Table~\ref{table:ala2}.
%
%The convergence analysis is done in the {\color{red} \tt SI~Section~XX}.
%TODO: Convergence of free energy barriers as a function of time for WTMTD, TASS-WHAM, TASS-MF(40,30,20).
%
%TODO: Also make a plot of how many times configurations A and B are visitied in the span of every 5~ns 
%in each of these simulations
%
%
In the WTMTD simulations, free energy barriers \rxn{A}{B} and \rxn{B}{A} were converged to 9.7 and 8.0 kcal~mol$^{-1}$, respectively; See Appendix B for convergence study.
The same computed from TASS-WHAM converged to 9.4 and 7.8 %TODO: It has to be 7.4
kcal~mol$^{-1}$ in 20~ns per window; see also Appendix B. 
The differences between the free energy barriers from the two simulations are less than 1~kcal~mol$^{-1}$.
%, and is likely to be due to {\tt \color{red} WHAT??}.
%
Of great importance, the barriers computed from TASS-MF (40) with the same TASS windows agree well
with the TASS-WHAM (40) data.
The errors in the free energy barriers using TASS-MF (30), and TASS-MF (20) are also less than 1~kcal~mol$^{-1}$.
%,
Further, the $L^2$~errors (Appendix A), computed by taking the WTMTD free energy surface as the reference were found to be less than 0.6~kcal~mol$^{-1}$.
%while with 20 windows, it is about 2~kcal~mol$^{-1}$.
%
It is to be noted that WHAM will not converge well with 30 and 20 umbrella windows.
%

%The free energy barrier for ${\bf XX}\rightarrow{\bf YY}$ %TODO
%computed using WTMTD is XX~kcal/mol.
%
%As it can be seen from the Table~XX %TODO
%the free energy barrier computed using WTMTD is in %excellent agreement
%with TASS-MF results with 20, 30 and 40 windows. 
%
%TODO: Need a table (ROTATE)
%                       
%            | WTMTD |  TASS-WHAM-40 | TASS-WHAM-30 | TASS-MF-40 | TASS-MF-30 | TASS-MF-20
%  Delta_FXY |
%  Delta_FXY^dagger|
%            | WTMTD |  TASS-WHAM-40 | TASS-WHAM-30 | TASS-MF-40 | TASS-MF-30 | TASS-MF-20
%
 \begin{table}[ht]
    \centering
\begin{tabular}{ |c|c|c|c|c|c|c|c|c| }
\hline \hline 
    \multirow{2}{*}{Method}      & \multicolumn{2}{c |}{$\Delta F^\ddagger$ (kcal~mol$^{-1}$)}  & \multirow{2}{*}{$L^2 {\rm error}$ (kcal~mol$^{-1}$)} \\ \cline{2-3} 
            & \rxn{A}{B} & \rxn{B}{A} &
             \\ \hline 
WTMTD            & 9.7 &  8.0 & 0.0 \\ \hline 
TASS-WHAM        & 9.4 &  7.8 & 0.4 \\\hline
TASS-MF (40)     & 9.7 &  7.9 & 0.5 \\ \hline
TASS-MF (30)     & 9.6 &  7.8 & 0.4 \\ \hline
TASS-MF (20)     & 8.9 &  7.1 & 0.6 \\  \hline \hline
\end{tabular}
\caption{Free energy barriers ($\Delta F^\ddagger$) 
for the transitions \rxn{A}{B} and \rxn{B}{A} computed from various methods. $L^2 {\rm error}$ was calculated by taking WTMTD as reference.}
\label{table:ala2}
\end{table}
\subsection{Alanine tripeptide}
\begin{figure}[htbp]
\centering
    \includegraphics[width=0.50\textwidth]{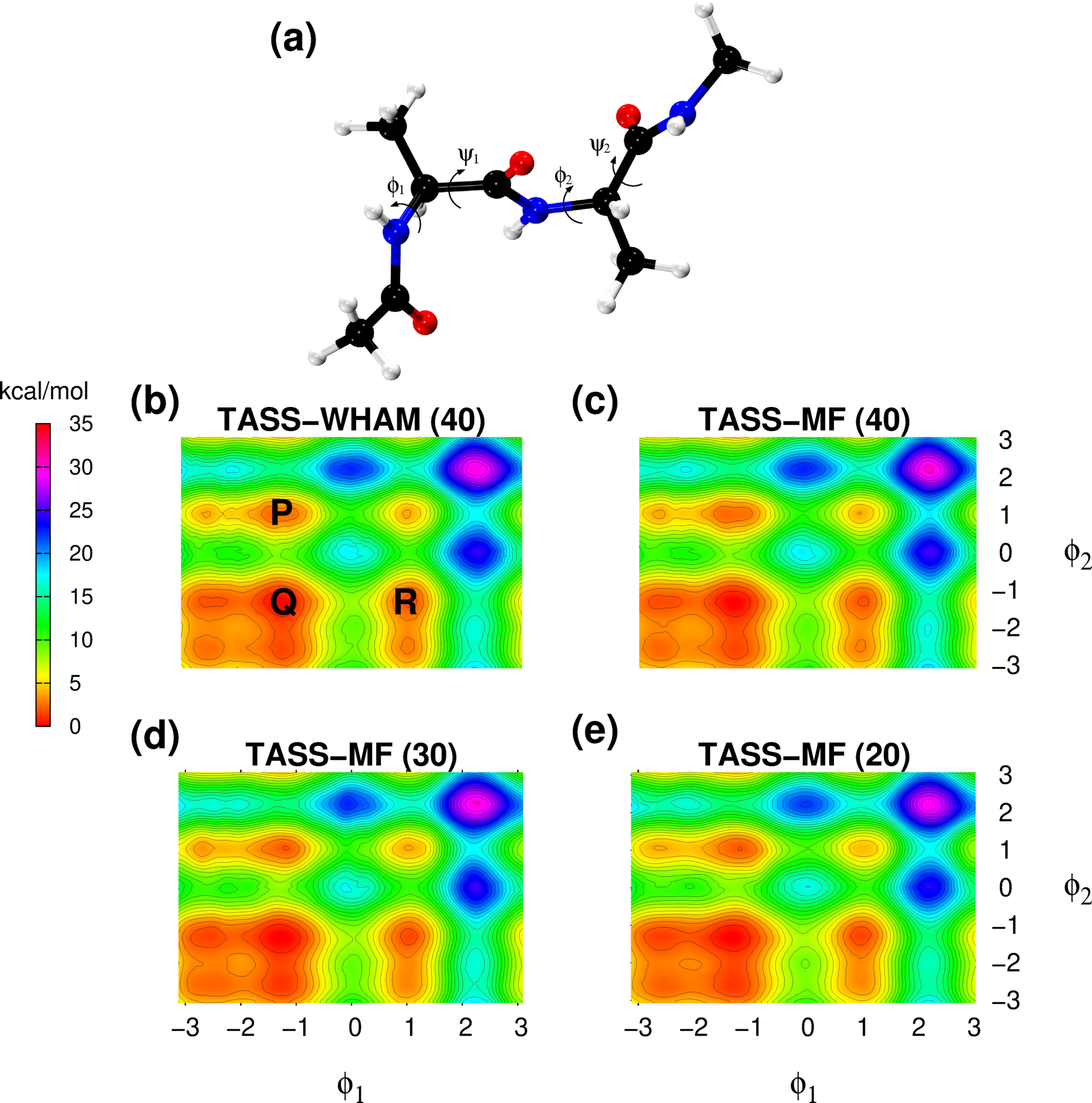}
    \caption{%{\tt \color{red} FIGURE 1 of paper}
    (a) Ball and stick representation of alanine
tripeptide.
%$\phi$ and $\psi$ are the Ramachandran angles, specifically the torsional angle between atoms C-N-C$_{\alpha}$-C and N-C$_{\alpha}$-C-N, respectively, as shown in the figure. Color code: H (white), C (black), O(red), and N (blue). {\tt{\color{red} color code is not required here as it is mentioned in alanine dipeptide figure.}} 
The free energy surface reconstructed in the $(\phi_1,\phi_2)$-space computed from (b) WHAM-40 can be compared with those using the TASS-MF method with 40 (c), 30 (d), and 20 (e) equidistant
windows. The contours are drawn for every 1 kcal mol$^{-1}$.}
    \label{fig:alatri_phi1_phi2}
\end{figure}

\begin{figure}[htbp]
\centering
    \includegraphics[width=0.50\textwidth]{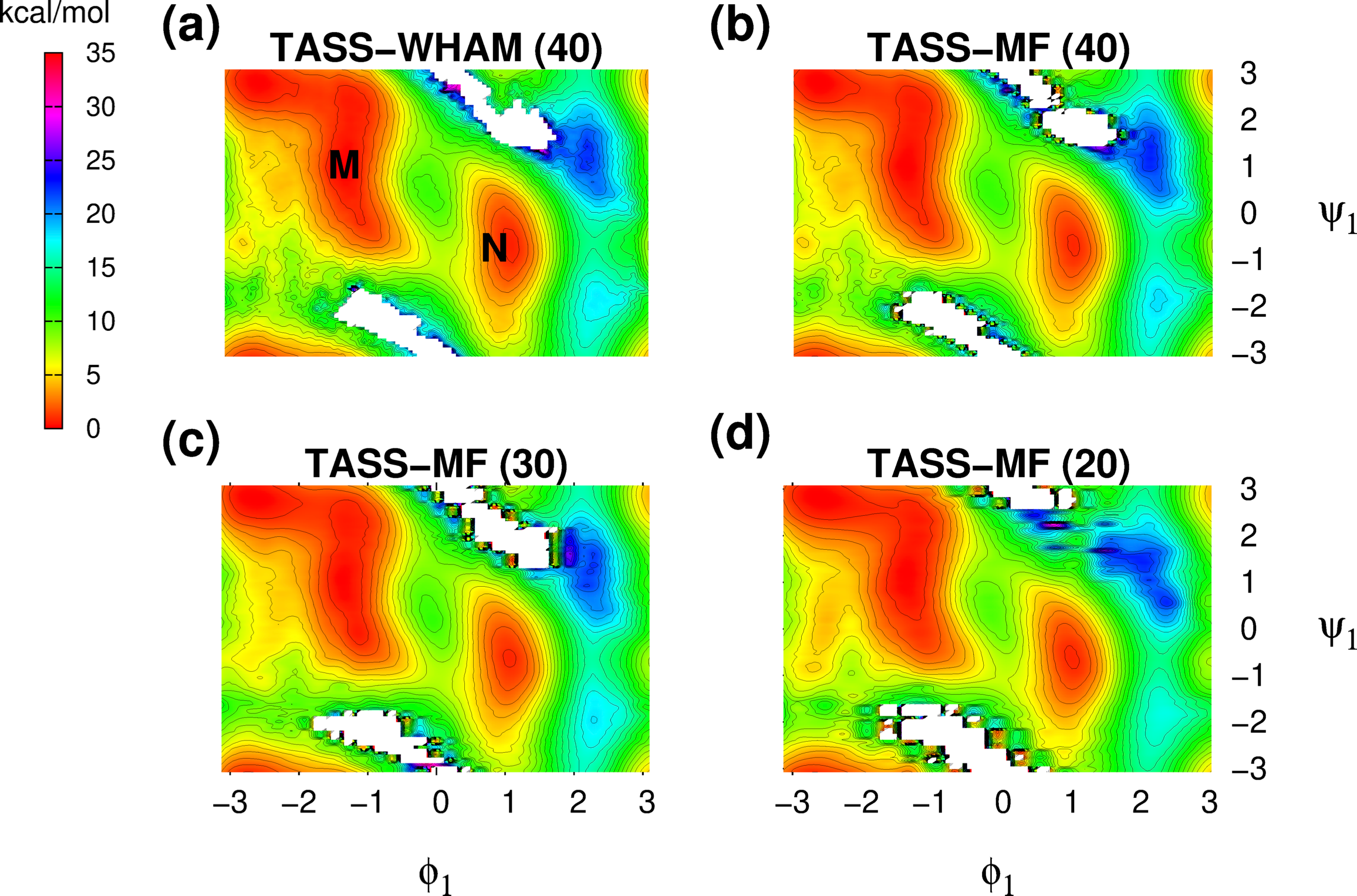}
    \caption{%{\tt \color{red} FIGURE 1 of paper}
    (a) The free energy surfaces  $F(\phi_1,\psi_1)$  computed from TASS-WHAM with 40 equidistant windows is presented together with the same computed using TASS-MF with 40 (b), 30 (c) and 20 (d)  equidistant
windows. The contours are drawn for every 1 kcal mol$^{-1}$.}
    \label{fig:alatri_phi1_psi1}
\end{figure}

\begin{figure}[htbp]
\centering
    \includegraphics[width=0.50\textwidth]{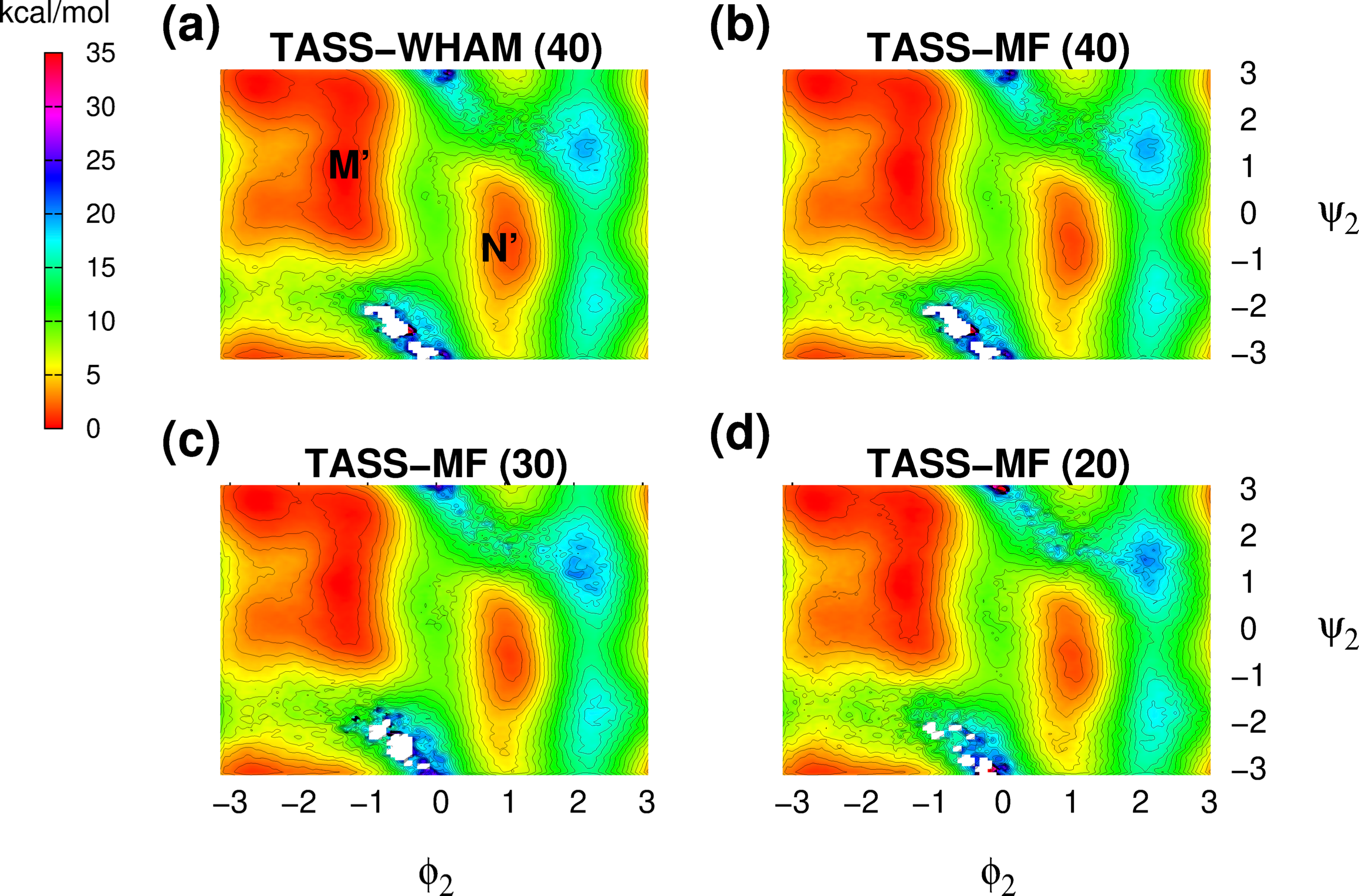}
    \caption{%{\tt \color{red} FIGURE 1 of paper}
(a) The free energy surfaces  $F(\phi_2,\psi_2)$  computed from TASS-WHAM with 40 equidistant windows is presented together with the same computed using TASS-MF with 40 (b), 30 (c) and 20 (d)  equidistant
windows. The contours are drawn for every 1 kcal mol$^{-1}$.}
    \label{fig:alatri_phi2_psi2}
\end{figure}

To demonstrate the application of the method to high-dimensional free energy landscapes, we reconstructed the four dimensional free energy landscape of alanine tripeptide \textit{in vacuo} as a function of four Ramachandran angles $(\phi_1,\psi_1,\phi_2,\psi_2)$; see Fig.~\ref{fig:alatri_phi1_phi2}(a).
We note that the four-dimensional free energy surface was available in our calculation from TASS but not from WTMTD, since the latter was computationally prohibitive.
Here, umbrella bias was applied along $\phi_1$, metadynamics bias was applied along $\phi_2$, while all the four CVs were enhanced sampled at high temperature as per the TASS formalism.
Equilibration run for each window was carried out for about 1~ns followed by 40~ns of production run.
All the parameters and simulation details for the TASS simulation were kept the same as in the case of alanine dipeptide except that $\mu$ was taken as  $450$~amu {\AA}$^2$ rad$^{-2}$.
After carrying out the TASS simulation with four CVs $(\phi_1,\psi_1,\phi_2,\psi_2)$ the reconstructed four-dimensional free energy was projected along $(\phi_1,\phi_2)$, $(\phi_1,\psi_1)$ and $(\phi_2,\psi_2)$ spaces; see Figs.~\ref{fig:alatri_phi1_phi2}(b)-(e), \ref{fig:alatri_phi1_psi1}, and \ref{fig:alatri_phi2_psi2}.
The performance of TASS-WHAM with 40 umbrella windows was compared with TASS-MF with 40, 30, and 20 windows.
%
%Similar to the alanine dipeptide case, TASS-MF was carried out with 20, 30, and 40 umbrella windows here as well.
% 
%Spline interpolation was used to obtained smooth projected surfaces.
%F(\phi_1,\phi_2)$ and  $F(\phi_1,\psi_1)$ surface of MF reweighting.
%
%$F(\phi_2,\psi_2)$ surface was generated by 2-dimensional projection of $F(\phi_1,\phi_2,\psi_2)$.
%
The comparison of the free energy barriers
on the $F(\phi_1,\phi_2)$, $F(\phi_1,\psi_1)$, and $F(\phi_2,\psi_2)$ surfaces, computed from TASS-WHAM and TASS-MF simulations,
are presented in Tables~\ref{table:ala3_phi1_phi2}.
%{\tt \color{red} is it by WHAM?} 
%
%Several major minima can be found in these figures. 
%
The difference in the free energy barriers computed from TASS-WHAM (40) and TASS-MF (40) is not greater than 0.2~kcal~mol$^{-1}$.
Free energy barriers computed from TASS-MF using 40, 30, and 20 windows are agreeing within an acceptable margin of 1 kcal~mol$^{-1}$.
The $L^2$ errors in the $F(\phi_1,\phi_2)$ estimates by the TASS-MF calculations (by taking TASS-WHAM as the reference) are also less than or equal to 0.5~kcal~mol$^{-1}$, and the highest error was observed for TASS-MF (20).
\begin{table*}[btp]
    \centering
\begin{tabular}{ |c|c|c|c|c|c|c|c|c|c| }
\hline \hline 
\multirow{2}{*}{Method}  & \multicolumn{8}{c |}{$\Delta F^\ddagger$ (kcal~mol$^{-1}$)}  & \multirow{2}{*}{$L^2 {\rm error}$ (kcal~mol$^{-1}$) } \\ \cline{2-9}
            & \rxn{P}{Q} & \rxn{Q}{P} & \rxn{R}{Q} & \rxn{Q}{R} 
            & \rxn{M}{N} & \rxn{N}{M} & \rxn{M$^\prime$}{N$^\prime$} & \rxn{N$^\prime$}{M$^\prime$} & \\ \hline
TASS-WHAM (40)   & 6.6 & 8.9 & 6.7 & 8.5 & 7.4 & 6.6 & 9.1 & 7.7 & 0.0 \\\hline
TASS-MF (40)     & 6.6 & 8.9 & 6.7 & 8.3 & 7.2 & 6.6 & 9.0 & 7.6 & 0.1 \\ \hline
TASS-MF (30)     & 6.7 & 8.7 & 6.5 & 8.1 & 7.5 & 6.7 & 9.0 & 7.8 & 0.4 \\ \hline
TASS-MF (20)     & 6.7 & 8.7 & 6.3 & 7.8 & 6.9 & 6.1 & 8.9 & 7.4 & 0.5 \\  \hline \hline
\end{tabular}
\caption{Free energies barriers ($\Delta F^\ddagger$) 
for \rxn{P}{Q}, \rxn{Q}{P}, \rxn{R}{Q}, and \rxn{Q}{R} in $(\phi_1,\phi_2)$ space, \rxn{M}{N}, \rxn{N}{M} in ($\phi_1$,$\psi_1$) space, \rxn{M$^\prime$}{N$^\prime$}, and \rxn{N$^\prime$}{M$^\prime$} in ($\phi_2$,$\psi_2$) space computed from various methods.  $L^2 {\rm error}$ was calculated for $F(\phi_1,\phi_2)$ by taking TASS-WHAM (40) as reference.}
\label{table:ala3_phi1_phi2}
\end{table*}

\section{\label{sec:concl}Conclusions}
An efficient as well as computationally straightforward reweighing scheme for TASS is presented here.
In this approach the high-dimensional free energy estimate is divided into two parts, in which the first part contains the free energy projected along the umbrella sampling coordinate, while the second term 
is the low-dimensional slice of the free energy surface.
We show that the first term can be directly computed by integrating the reweighted mean force acting along the umbrella coordinate at the equilibrium position of the umbrella potential, whereas the second term can be obtained by binning and reweighting.
%
%For every umbrella, the two terms can be computed simultaneously and then free energy can be obtained by adding these two terms.  
%We have shown that our reweighing procedure is capable of reproducing the complex high dimensional free energy surfaces for folding of oligopeptides like dialanine and trialanine obtained through our previously reported TASS reweighing scheme.
The advantage of the new method is that WHAM iteration can be completely avoided in order to combine free energy slices. 
We demonstrated the accuracy of the method by reconstructing the free energy surfaces of alanine dipeptide and alanine tripeptide systems.
The surfaces were computed accurately even with half the number of umbrella windows, with a maximum difference in free energy barriers not beyond than 0.5~kcal~mol$^{-1}$ compared to that of TASS-WHAM.
As the TASS-MF method proposed here permits us to perform reconstruction of high-dimensional free energy surfaces with much less number of umbrella windows, the method improves the efficiency of the TASS simulations.
%
%shown that unlike the previous reweighing technique where WHAM has a serious limitation for computing free energies using less umbrella windows, the new method is efficient in computing the free energy even if less number of umbrella windows are used and thus make the TASS effective for all ranges of umbrellas which further reduces the computational cost and complexity of the post processing of TASS.
 %\pagebreak
 %\newpage
 %\clearpage
%
\begin{acknowledgments}
Authors acknowledge the HPC facility at the Indian Institute of Technology Kanpur (IITK) for the computational resources. 
NN is grateful to the Institute of Catalysis, Hokkaido University (Japan) for awarding the Visiting Professorship. This project was started during his stay at the Hokkaido University.
NN is indebted to Prof. Akira Nakayama for the valuable discussions and inputs.
SV acknowledges DST-INSPIRE for the PhD fellowship.
MS thanks financial support from JSPS KAKENHI (18H05519, 18K05208, 21H01603) and MEXT Program for Promoting Researches on the Supercomputer Fugaku (Fugaku Battery and Fuel Cell Project).
%Computations were performed using the HPC cluster at Indian Institute of Technology Kanpur.
%
\end{acknowledgments}
\appendix
\section{Calculation of Least Square Error: $L^2 \mathrm{error}$ \cite{Tuckerman:2012}}
\begin{equation}
    L^2 \mathrm{error} = \sqrt{\frac{1}{N} \sum_{i}^{N} [F(\mathbf{z}_i) - F_\mathrm{ref}(\mathbf{z}_i)]^2 }
\end{equation}
Here, $N$ is the total number of the grid points, $F(\mathbf{z}_i)$ is the value of computed free energy for various methods and $F_\mathrm{ref}(\mathbf{z}_i)$ is the value of reference free energy at $i^\mathrm{th}$ grid.
\section{Convergence of Free Energy Barriers: Alanine Dipeptide In Vaccuo}
\begin{figure}[htbp]
\centering
    \includegraphics[width=0.40\textwidth]{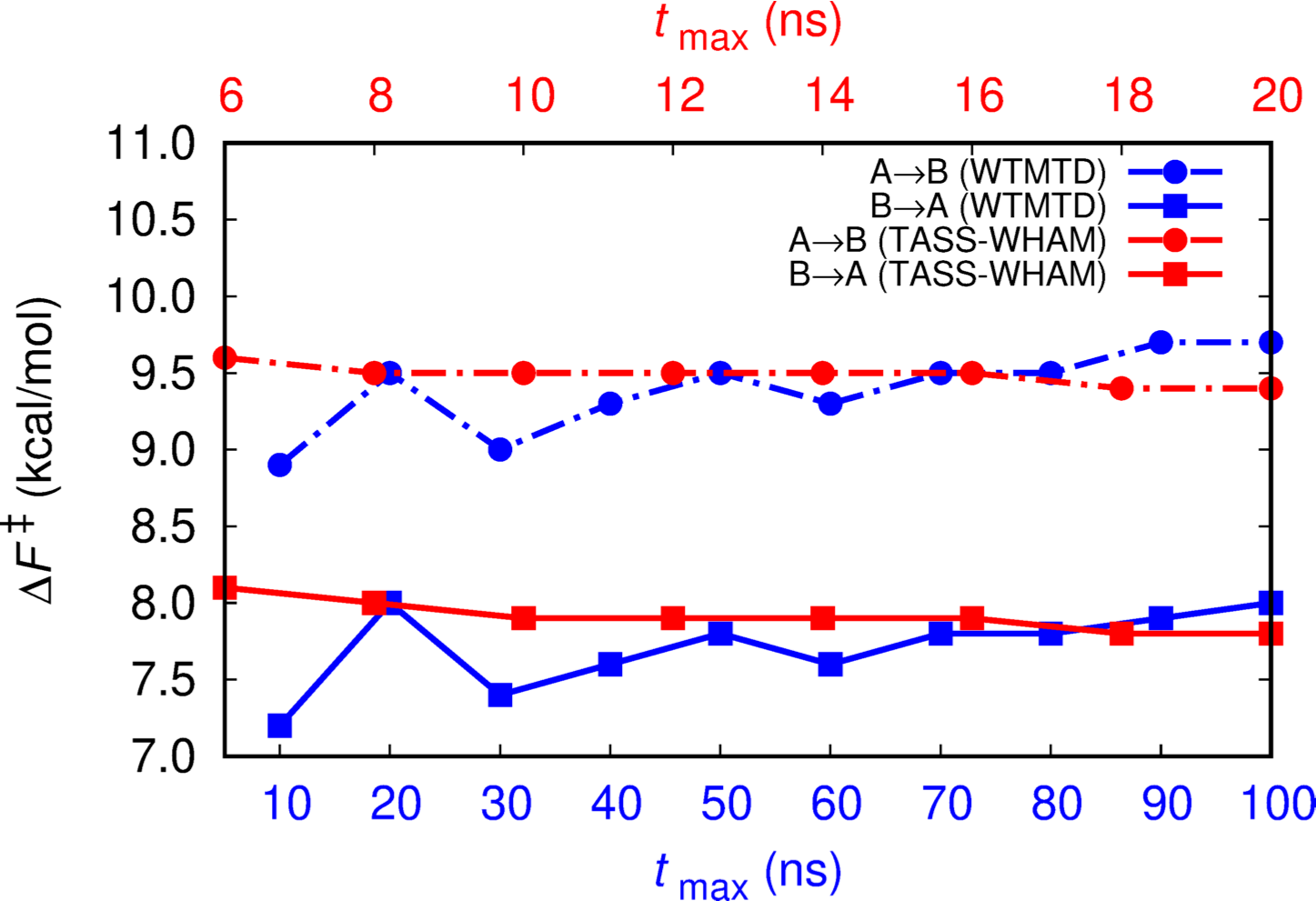}
    \caption{Convergence of free energy barriers ($\Delta F^ \ddagger$) for \rxn{A}{B} in WTMTD (blue circles) and TASS-WHAM (red circles) simulations is plotted together with that for \rxn{B}{A} (blue/red squares). Here, $t_{\mathrm{max}}$ is the net simulation length in WTMTD and simulation time per umbrella window for TASS-WHAM (40 windows).
    }
    \label{fig:aladi_conv}
\end{figure}
%
%%%%%%%%%%%%%%%%%%%%%%%%%%%%%%%%%%%%%%%%%%%%%%%%%%%%%%%%%%%%%%%%%%%%%%%%%%%%%%%%%%%%%
%\nocite{*}
\bibliography{references.bib}% Produces the bibliography via BibTeX.

\end{document}